\documentclass[12pt]{article}
\usepackage{mainstyle}
\usepackage[backend=bibtex, maxnames=4, sorting=none, citestyle=numeric-comp, bibstyle=trad-unsrt]{biblatex}
\usepackage[a4paper,
            left=2cm,
            right=2cm,
            top=2cm,
            bottom=2cm]{geometry}

\usepackage{placeins}
\usepackage{subfig}

\title{Optimization and robustness of cost-efficient\\ seismic arrays for Newtonian noise cancellation\\ at the Einstein Telescope}

\author{Patrick Schillings\footnote{patrick.schillings@rwth-aachen.de}\, and Johannes Erdmann}

\date{\small RWTH Aachen University, III. Physikalisches Institut A, Aachen, Germany}

\begin{document}

\maketitle

\begin{abstract}
\noindent
Newtonian noise is expected to be the dominating noise source for low frequencies at the Einstein Telescope.
It originates from seismic waves that cause density fluctuations in the rock around the interferometer.
The mitigation strategy for Newtonian noise relies on an array of seismometers, placed at depth in boreholes, which provides measurements of the seismic wave field.
We optimize the positions of the individual seismometers for the mitigation capabilities of the array for a full corner of the Einstein Telescope.
We find that the mitigation capabilities of arrays with multiple seismometers in each borehole match the capabilities of only somewhat smaller arrays but with only one seismometer per borehole.
Mitigation is further improved by extending the array with seismometers in the interferometer tunnels.
Such configurations may hence provide a cost-effective way towards realizing an efficient seismic array.
In each case, we quantify the broadband mitigation performance in the range from 1 to $\SI{10}{\Hz}$ for arrays that are optimized for a frequency of $\SI{10}{\Hz}$, as well as the robustness of the arrays with respect to variations from their optimized positions.
We find that larger arrays with several seismometers per borehole and additional seismometers in the tunnels provide promising broadband performance above 3 to $\SI{4}{\Hz}$ and that such arrays are particularly stable against variations in the seismometer positions with mitigation factors $>6$ for an array of 20 boreholes with 3 seismometers each and $>15$ for a large array of 50 boreholes with 10 seismometers. 
\end{abstract}

\section{Introduction}
\label{sec:Intro}
The Einstein Telescope (ET)~\cite{ETDesignReportUpdate} is a proposed third generation, underground gravitational wave interferometer. Its aim is to significantly increase the sensitivity compared to the current gravitational wave interferometers LIGO~\cite{LIGOpaper}, Virgo~\cite{Virgopaper} and Kagra~\cite{KAGRApaper,KAGRApaper2}, where a key design goal is to enable measurements of gravitational waves down to frequencies of $\SI{3}{\Hz}$. The limiting noise source for ET's low frequency interferometer is expected to be Newtonian noise~\cite{NewtonianNoiseOrigin}. It arises from the gravitational force of surrounding density fluctuations that are caused predominantly by seismic body waves~\cite{SiteSelectionCriteria}. Seismic P-waves are longitudinal and directly compress and dilute the rock, which results in a gravitational pull on the interferometer mirrors~\cite{terrestialGravityFluctuations}. 
Seismic S-waves are transversal and plane S-waves do not cause direct density fluctuations. But both kinds of waves additionally contribute to Newtonian noise by shifting volumes of different densities, most importantly, the cavern walls~\cite{terrestialGravityFluctuations}. To mitigate Newtonian noise, it has been proposed to implement an array of auxiliary sensors that measure the seismic wave field and from which Newtonian noise can be predicted in order to subtract it from the interferometer output~\cite{Cella}. Since boreholes need to be drilled to place the auxiliary sensors, e.g.\ seismometers, in the rock around the underground ET, which comes at considerable costs, the positions of these sensors must be carefully optimized with respect to their ability to predict Newtonian noise. First efforts for such an optimization were made in Ref.~\cite{FrancescaSingleMirrorOptimization} for a single interferometer mirror and then expanded to include a full corner of four correlated mirrors~\cite{FrancescaJointMirrorOptimization}. An improved optimization algorithm was developed in Ref.~\cite{OurPaper}, which enables the efficient optimization of larger arrays. As a test bed for Newtonian noise cancellation, an optimized seismometer array was deployed at the Virgo interferometer~\cite{ArrayOptimizationForVirgo}, where a tiltmeter was used as a proxy for Newtonian noise and successfully predicted~\cite{VirgoNewtonianNoiseCancellingSystem}.

Different algorithms were studied for estimating Newtonian noise from the auxiliary seismometer measurements: stationary~\cite{terrestialGravityFluctuations,FrancescaSingleMirrorOptimization,FrancescaJointMirrorOptimization,OurPaper,LIGO_NNreductionSimulation,ArrayOptimizationForVirgo,VirgoNewtonianNoiseCancellingSystem,ophardt2025silencing,rading2025distributed,vanBeveren:2023seq} and adaptive Wiener filters~\cite{VirgoNewtonianNoiseCancellingSystemPart2}, as well as neural networks~\cite{vanBeveren:2023seq,NNNN}.
Recently, the potential of fusion arrays, made of seismometers and strainmeters, was studied for Newtonian noise mitigation~\cite{ophardt2025silencing}.
Here, we limit ourselves to the stationary case, a seismic array with seismometers only, and mitigation with stationary Wiener filters.

A key finding of Ref.~\cite{OurPaper} was that an array with a larger number of seismometers promises significantly improved prospects for Newtonian noise mitigation. 
We extend these studies by testing the performance of optimized arrays for frequencies in the range from 1 to $\SI{10}{\Hz}$, and we also test the robustness of these seismic arrays with respect to random displacements of the seismometers from their optimized positions.
In addition, we study array optimization under spatial constraints, where we require several seismometers to be in the same borehole. 
As the costs for the seismic array are largely dominated by the costs for the drilling of the boreholes, this may point to a cost-effective way to improve its mitigation capabilities for Newtonian noise. 
In addition, we add seismometers within the ET infrastructure, i.e., in the tunnels of the interferometer arms, which was found to be very effective for strainmeters~\cite{ophardt2025silencing}.
Also these seismometers come with significantly lower costs compared to those for additional boreholes.

We describe the methodology in Section~\ref{sec:MethodsAndScenarios} and discuss the results in Section~\ref{sec:Results}. We then conclude in Section~\ref{sec:Conclusions}.

\section{Methods and Scenarios}
\label{sec:MethodsAndScenarios}
The optimization method is described in detail in Ref.~\cite{OurPaper}. Here, we only give a brief outline:
The optimal positions are determined by minimizing the mean residual of four mirrors of a corner of the Einstein Telescope. A single mirror residual, $R_i$, is defined as 
\begin{equation}
    R_i(f)=\frac{\EW{\abs{s(f)-\tilde{s}(f)}^2}}{\EW{\abs{s(f)}^2}}=1-\frac{\vec{c}_\text{sd}^\dagger\cdot C_\text{dd}^{-1} \cdot \vec{c}_\text{sd}}{c_\text{ss}},
    \label{eq:residual}
\end{equation}
where $s(f)$ is the Newtonian noise signal in frequency space $f$, $\tilde{s}(f)$ is the signal estimation with a Wiener filter, $\EW{\cdot}$ and the expectation value. This is equal to a combination of power spectral densities (PSD), where $c_{\text{ss}}=\EW{\abs{s(f)}^2}$ is the signal PSD, $\vec{c}_\text{sd}=\EW{s^*\vec{d}}$ is the cross power spectral density (CPSD) of signal and data, and $C_\text{dd}=\EW{\vec{d}^*\vec{d}}$ is the data CPSD, where $\vec{d}(f)$ is the data vector that contains the data of all seismometer channels.

Under the assumption of a large homogeneous rock around a small spherical cavern and a stationary, isotropic, monochromatic plane wave field, the expression for $R_i(f)$ is computed analytically~\cite{FrancescaSingleMirrorOptimization} using the Wiener filter~\cite{WienerFilter} as the prediction algorithm. Additionally, the correlation of the close-by inner mirrors are taken into account~\cite{FrancescaJointMirrorOptimization}. To find the optimal positions, we first find good positions with the particle swarm optimization (PSO) algorithm~\cite{PSOAlgorithm} and then continue the optimization with the differential optimizer Adam~\cite{AdamAlgorithm}, following the proposal from Ref.~\cite{OurPaper}. With respect to Ref.~\cite{OurPaper}, we only slightly improve the settings of the optimization algorithms as follows: PSO swarm size 400 instead of 800, and Adam patience 200 instead of 500 with a tolerance of $10^{-4}$ instead of $10^{-3}$.

The mitigation factor of a given array of seismometers is then defined as 
\begin{equation}
    M(f) \equiv \frac{1}{\sqrt{\max_i(R_i(f))}}=\frac{1}{\sqrt{R}} \, .
\end{equation}
It is an indicator of how much the Newtonian noise amplitude can be mitigated, where we conservatively consider only the maximum residual among the four mirrors~\cite{FrancescaJointMirrorOptimization}, while the position optimization uses the mean residual of the mirrors, as building the mean is differentiable~\cite{OurPaper}.

We only study arrays that are optimized for a single frequency of $f_\text{opt}=\SI{10}{\Hz}$ and examine their performance for mitigation Newtonian noise over the full band of frequencies between $f=\SI{1}{\Hz}$ and $f=\SI{10}{\Hz}$. We call this \say{broadband} performance. We chose \SI{10}{\Hz} because we found that arrays that are optimized for larger frequencies generalize much better for lower frequencies than vice versa. We note, however, that a broadband optimization of the array~\cite{FrancescaSingleMirrorOptimization} may yield even better results.

We test the robustness of the array by varying the seismometer positions in all three spatial directions using random displacements around their optimized positions that are drawn from a Gaussian distribution with a width of $\sigma=\SI{50}{\m}$, as proposed in Refs.~\cite{FrancescaSingleMirrorOptimization,FrancescaJointMirrorOptimization}.
We conservatively consider the larger variation of $\SI{50}{\m}$ from Ref.~\cite{FrancescaSingleMirrorOptimization} instead of $\SI{30}{\m}$ used in Ref.~\cite{FrancescaJointMirrorOptimization}. These variations are intended to account for uncertainties in the deployment of the seismic array and potential non-stationary effects in the seismic wave field~\cite{FrancescaSingleMirrorOptimization}. If, e.g., these variations are interpreted in terms of a varying wave speed, these correspond to variations of $\sim2.5\%$ for P-waves at \SI{3}{\Hz} and variations as large as $\sim12.5\%$ for S-waves at \SI{10}{\Hz}. We hence consider these variations quite large, in particular for the larger frequencies.

In previous studies, the seismometer positions were optimized independently, which implies that for each seismometer a new borehole needs to be drilled. In addition to this benchmark case, we study the mitigation performance of optimized arrays where multiple seismometers are placed inside one borehole, as for example suggested in Ref.~\cite{Amann:2022pyq}. This promises to be cost-efficient, as the costs for boreholes equipped with seismometers are driven by the drilling. In this case, the spatial variations from the optimized positions in $x$- and $y$-direction are performed for all seismometers in one borehole together, while they are independently displaced in $z$-direction, i.e., within the borehole.

Placing additional seismometers inside the ET infrastructure, i.e., in the interferometer tunnels and the caverns, comes with comparably low costs.
For fusion arrays that consist of seismometers and strainmeters, it was shown that constraining the strainmeters to the ET infrastructure is very efficient for Newtonian noise mitigation~\cite{ophardt2025silencing}. Here, we quantify the impact on the mitigation performance by adding a seismometer every $\SI{100}{\m}$ inside both arms up to $\SI{5}{\km}$ resulting in a total of 100 tunnel seismometers per corner. The positions of these seismometers are not optimized but they are used for Newtonian noise mitigation.

Consistent with Refs.~\cite{FrancescaJointMirrorOptimization,OurPaper}, we assume a signal-to-noise ratio of $\text{SNR}=15$, where ``signal'' is the seismometer displacement and ``noise'' is the seismometer noise, a P-wave fraction of $p=0.2$, a P-wave velocity of $c_\text{P}=\SI{6000}{\m\per\s}$, and an S-wave velocity of $c_\text{S}=\SI{4000}{\m\per\s}$. The optimization volume is constrained to $x,y=\pm\SI{1000}{\m}$ and $z=\pm\SI{300}{\m}$. P- and S-waves are treated as uncorrelated, the broadband seismometers have a flat response curve in the relevant frequency band and measure all three spatial directions, and we consider one corner of the triangle ET configuration with four mirrors (cf.~Table~1 of Ref.~\cite{OurPaper}).

\section{Results}
\label{sec:Results}

\begin{figure}
\makebox[\textwidth]{
    \centering    \subfloat{\includegraphics[width=.5\linewidth]{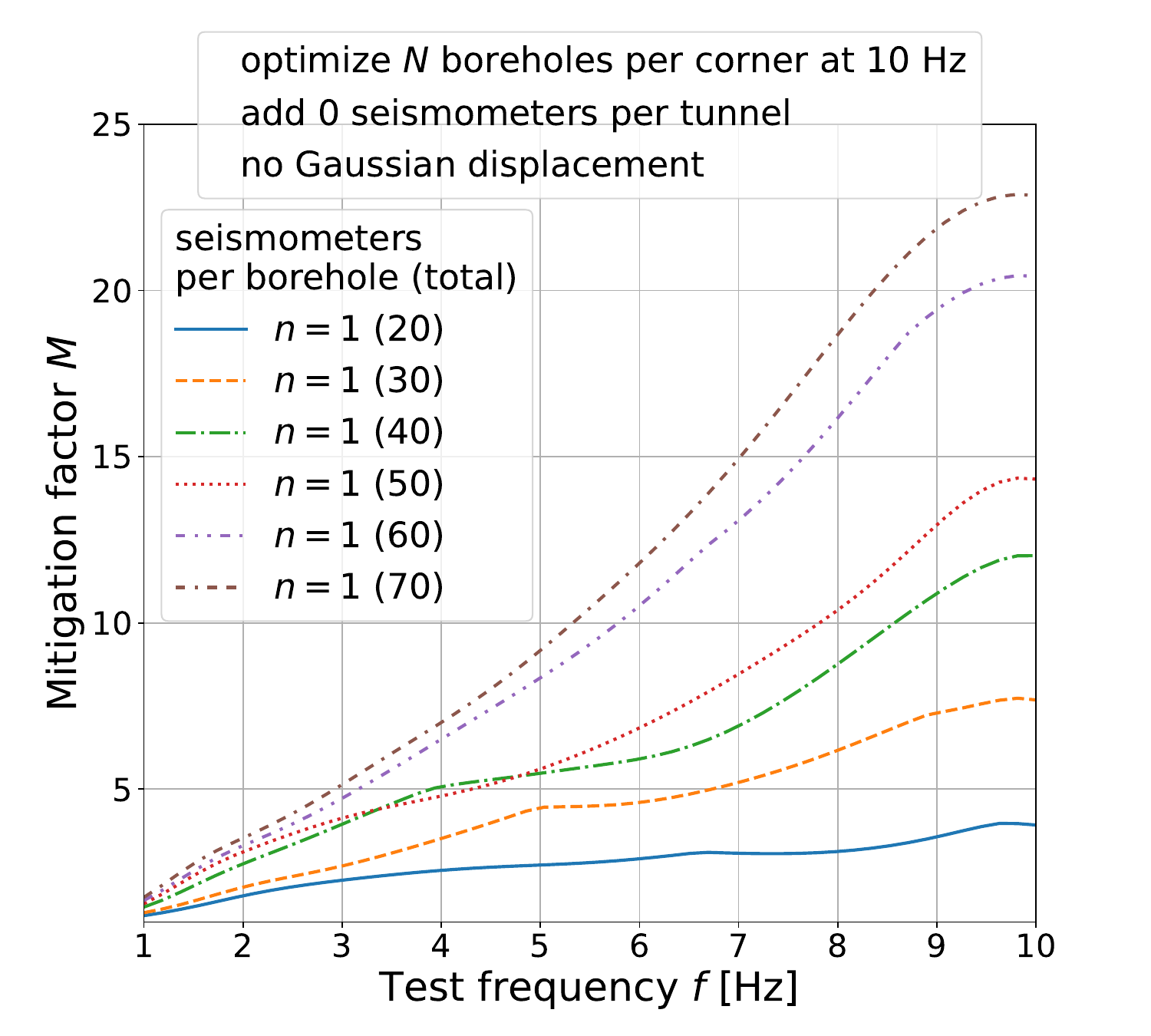}}
    \subfloat{\includegraphics[width=.5\linewidth]{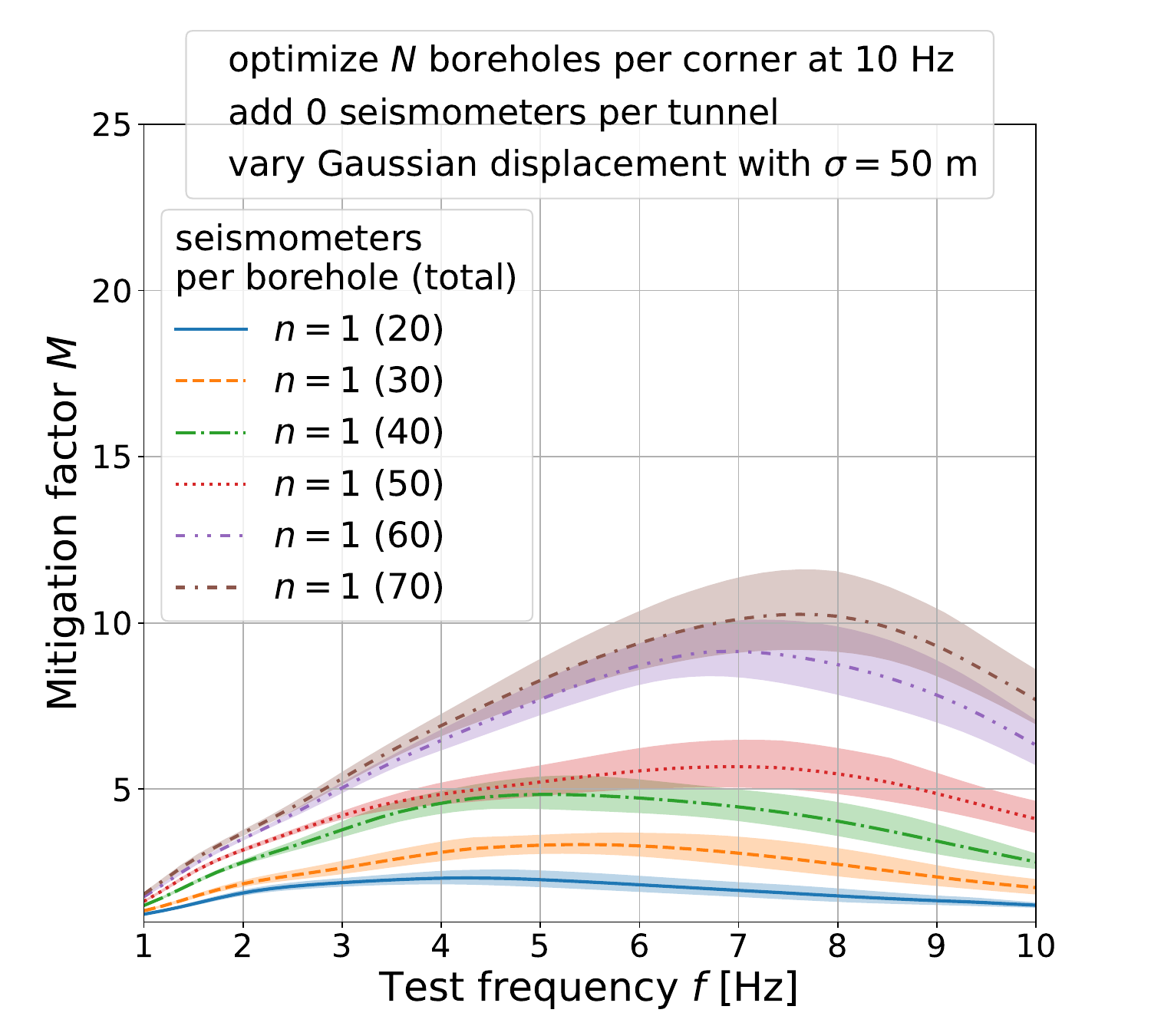}}
    }
    \caption{Left: Mitigation factor $M$ for Newtonian noise mitigation as a function of test frequency $f$ for different values of numbers of seismometers $N$ for $f_\text{opt}=\SI{10}{\Hz}$, $p=0.2$ and $\text{SNR}=15$. The optimization was performed with Adam after initialization with PSO. Right: Same, but in addition the optimized positions were displaced by a random distance drawn from a Gaussian distribution with $\sigma=\SI{50}{\m}$.}
    \label{fig:BroadbandN}
\end{figure}

\begin{figure}
\makebox[\textwidth]{
    \centering    \subfloat{\includegraphics[width=.5\linewidth]{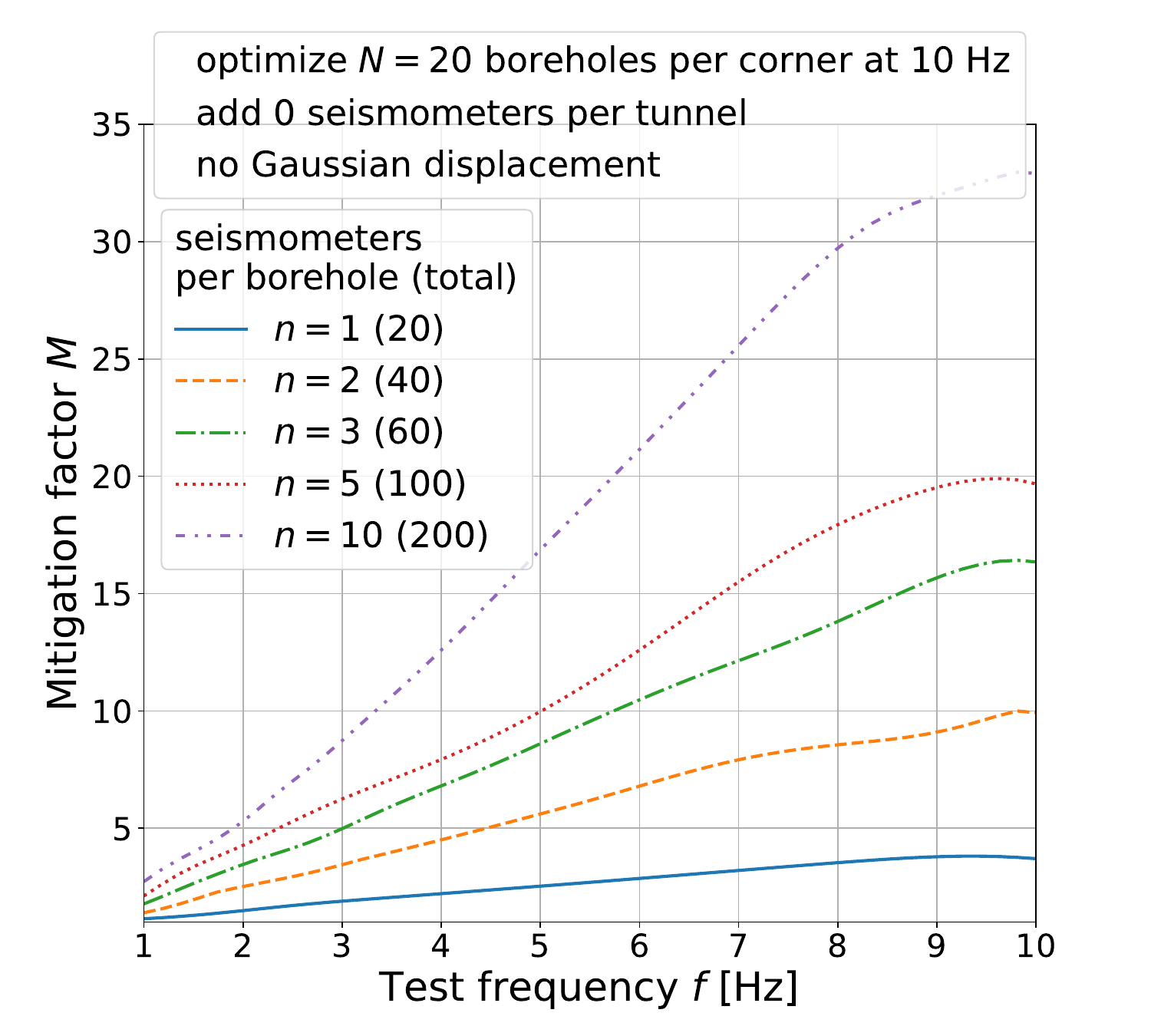}}
    \subfloat{\includegraphics[width=.5\linewidth]{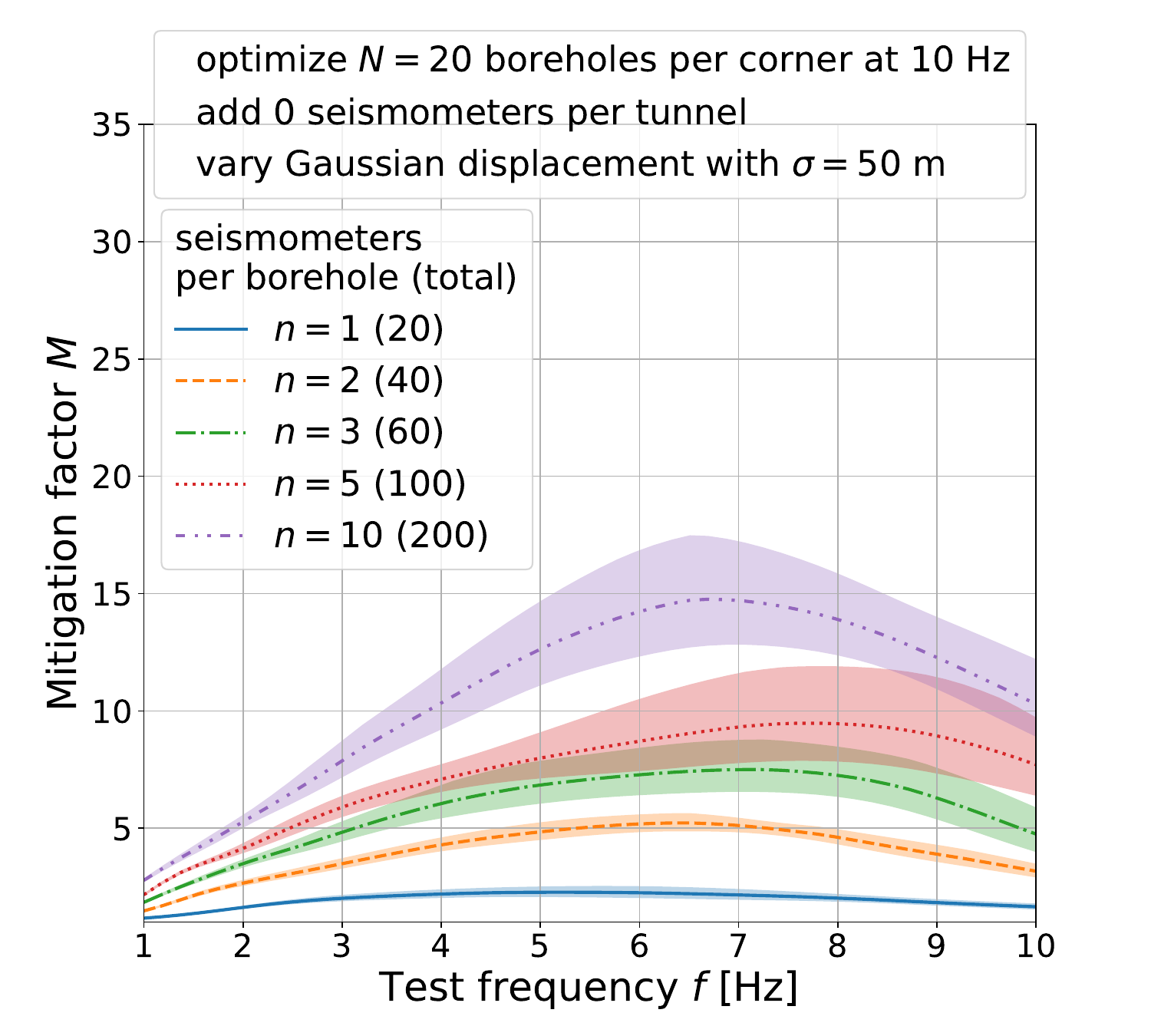}}
    }
    \caption{Left: Mitigation factor $M$ for Newtonian noise mitigation as a function of test frequency $f$ for different numbers of seismometers per borehole for $N = 20$ boreholes, $f_\text{opt}=\SI{10}{\Hz}$, $p=0.2$ and $\text{SNR}=15$. The optimization was performed with Adam after initialization with PSO. Right: Same, but in addition the optimized positions were displaced by a random distance drawn from a Gaussian distribution with $\sigma=\SI{50}{\m}$.}
    \label{fig:BroadbandMulti20}
\end{figure}

\begin{figure}
\makebox[\textwidth]{
    \centering    \subfloat{\includegraphics[width=.5\linewidth]{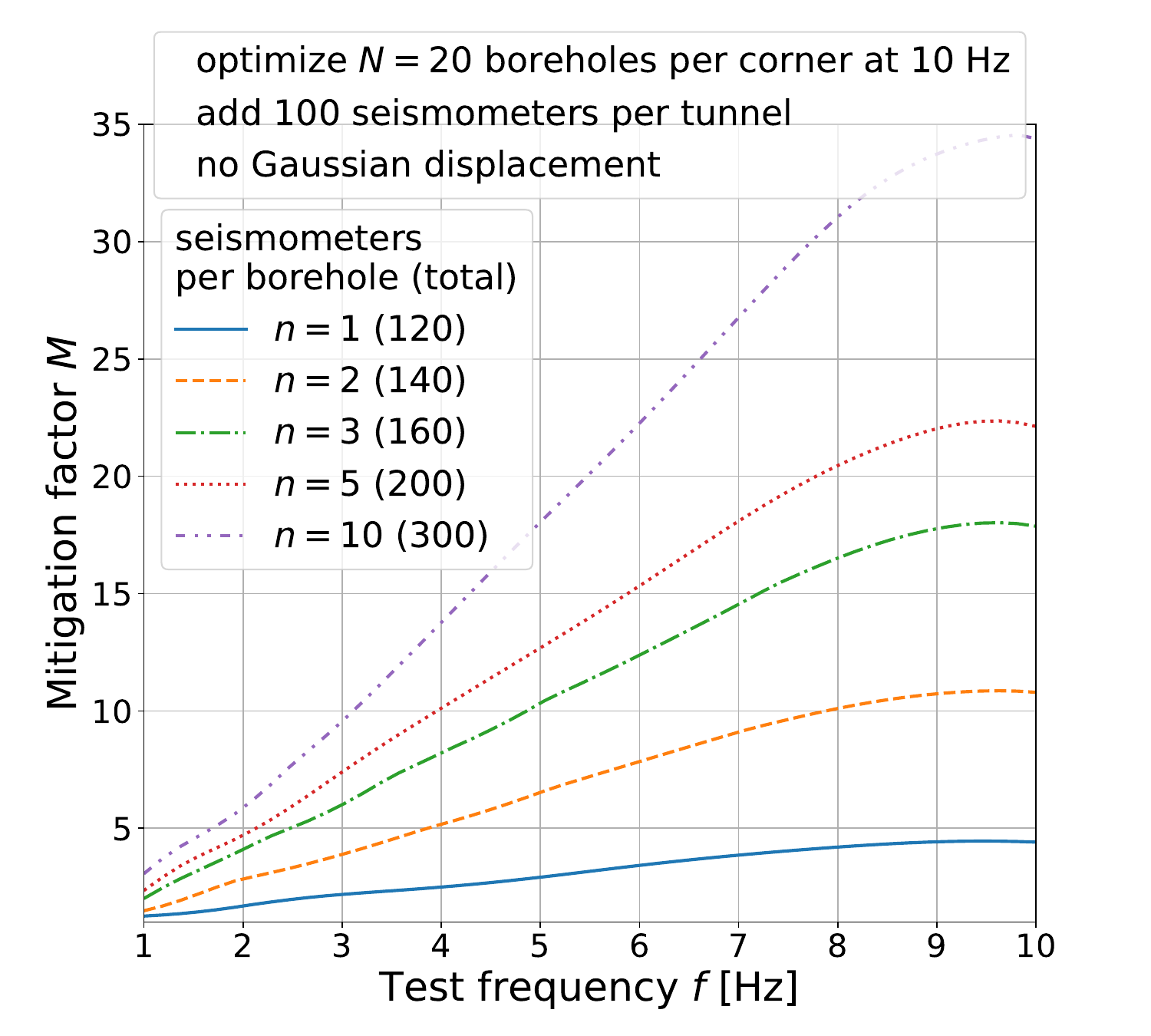}}
    \subfloat{\includegraphics[width=.5\linewidth]{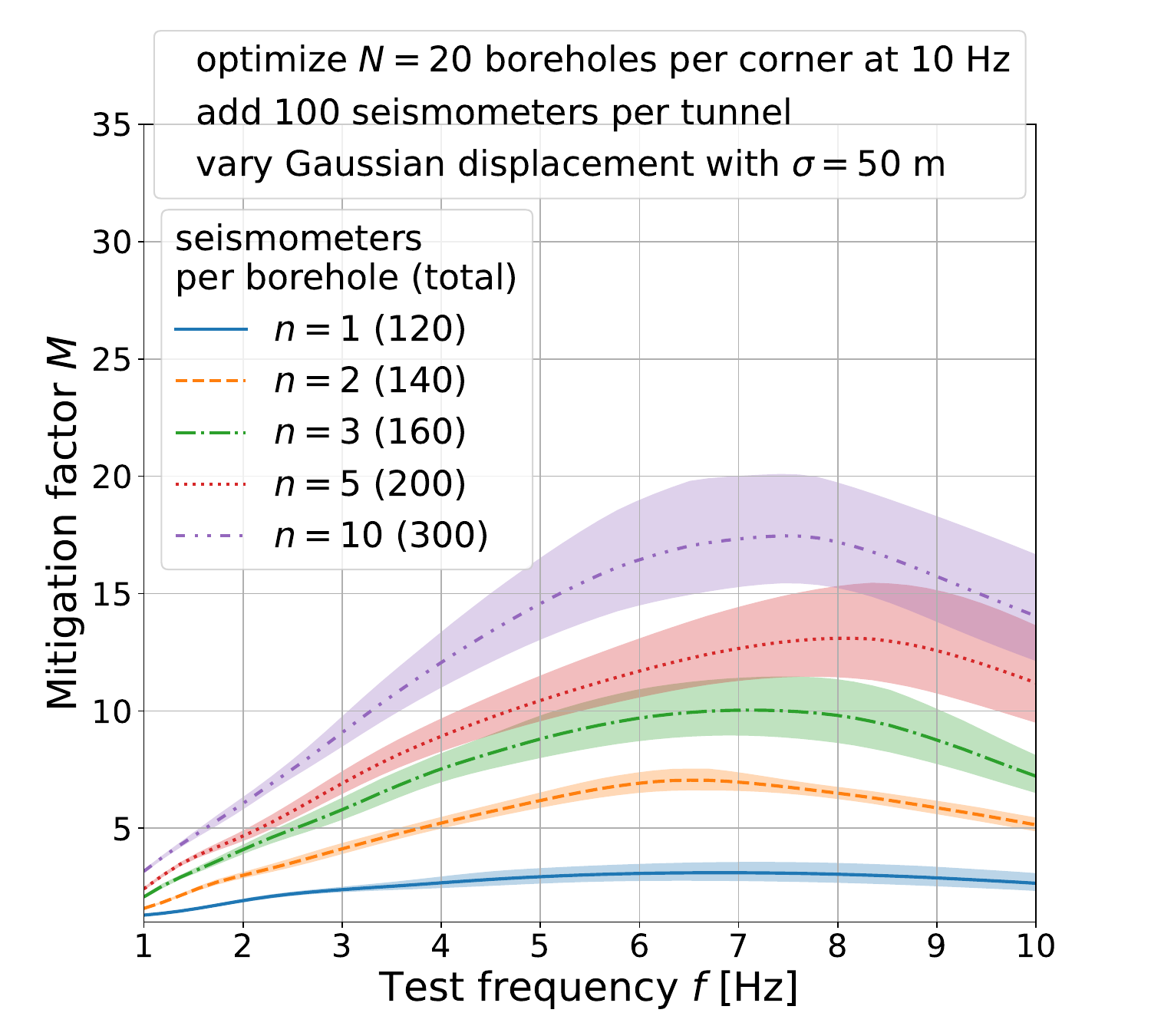}}
    }
    \caption{Left: Mitigation factor $M$ for Newtonian noise mitigation as a function of test frequency $f$ for different numbers of seismometers per borehole for $N = 20$ boreholes, $f_\text{opt}=\SI{10}{\Hz}$, $p=0.2$ and $\text{SNR}=15$. The optimization was performed with Adam after initialization with PSO. Additionally, 100 seismometers were added in the ET tunnels. Right: Same, but in addition the optimized positions were displaced by a random distance drawn from a Gaussian distribution with $\sigma=\SI{50}{\m}$.}
    \label{fig:BroadbandMulti20T100}
\end{figure}

\begin{figure}
\makebox[\textwidth]{
    \centering    \subfloat{\includegraphics[width=.5\linewidth]{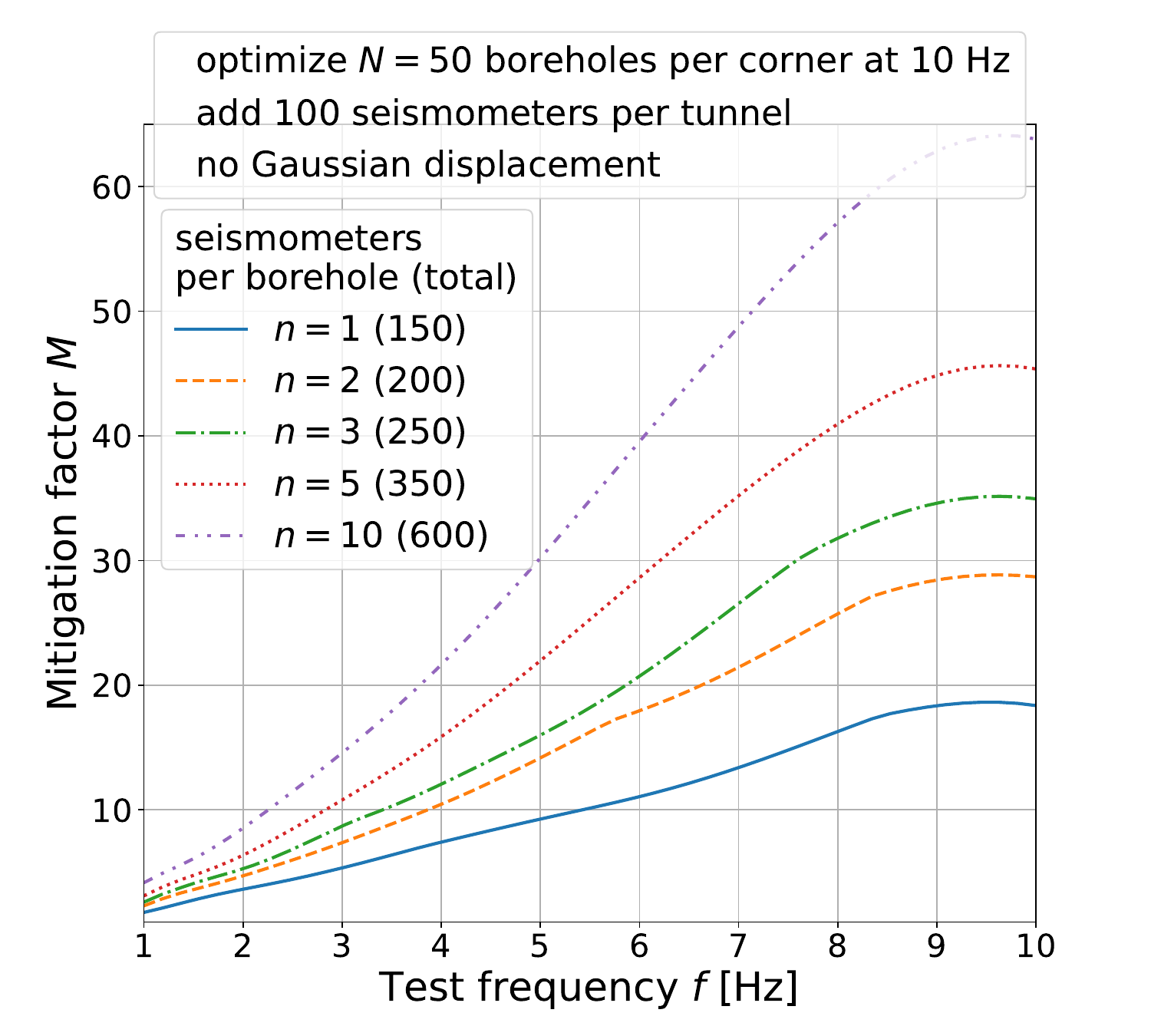}}
    \subfloat{\includegraphics[width=.5\linewidth]{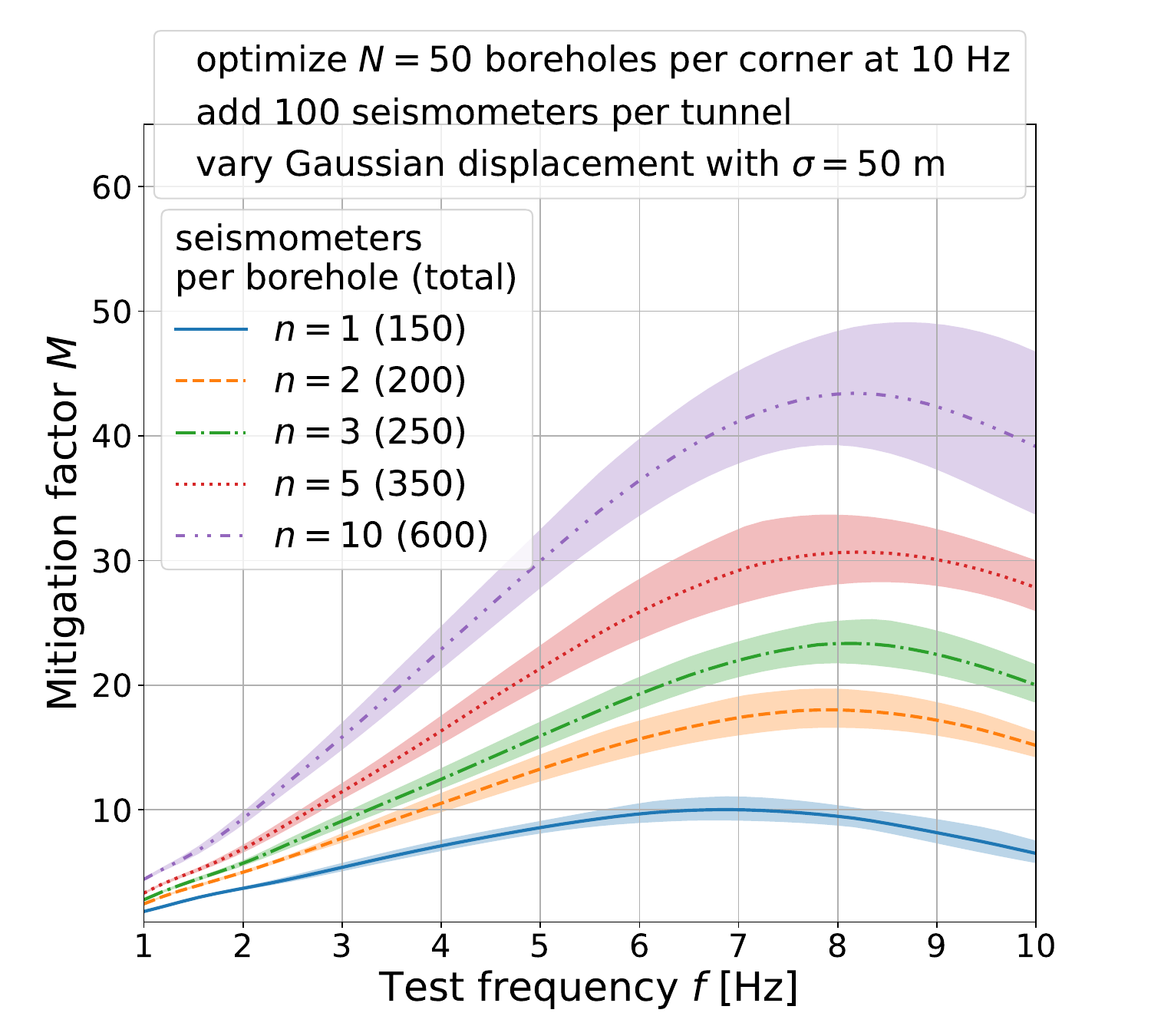}}
    }
    \caption{Left: Mitigation factor $M$ for Newtonian noise mitigation as a function of test frequency $f$ for different numbers of seismometers per borehole for $N = 50$ boreholes, $f_\text{opt}=\SI{10}{\Hz}$, $p=0.2$ and $\text{SNR}=15$. The optimization was performed with Adam after initialization with PSO. Additionally, 100 seismometers were added in the ET tunnels. Right: Same, but in addition the optimized positions were displaced by a random distance drawn from a Gaussian distribution with $\sigma=\SI{50}{\m}$.}
    \label{fig:BroadbandMulti50T100}
\end{figure}

We first present results on the broadband performance and robustness of arrays that are optimized with the setup from Ref.~\cite{OurPaper} (Fig.~\ref{fig:BroadbandN}). We then show results for 20 boreholes with a variable number of seismometers (Fig.~\ref{fig:BroadbandMulti20}) and with additional seismometers in the ET tunnels (Fig.~\ref{fig:BroadbandMulti20T100}), and finally for 50 boreholes with a variable number of seismometers and the additional seismometers in the tunnels (Fig.~\ref{fig:BroadbandMulti50T100}).

We start with discussing the broadband performance and robustness of arrays that are optimized in Ref.~\cite{OurPaper}: The array of seismometers is optimized with the Adam algorithm which was initialized with PSO at an optimization frequency of $f_\text{opt}=\SI{10}{\Hz}$. We consider only one seismometer per borehole and a total of 20, 30, \dots, 70 seismometers.


The broadband performance between $f=\SI{1}{\Hz}$ and $f=\SI{10}{\Hz}$ of the mitigation factor $M$ is shown in the left hand side of Figure~\ref{fig:BroadbandN}. The different curves show the performance of different numbers of seismometers $N$ from 20 to 70. The mitigation factor increases with the number of seismometers in the array because of the increased information about the wave field that is available to the Wiener filter. The best performance is always achieved for $f=\SI{10}{\Hz}$, which is the frequency for which the array was optimized. The seismometers of an optimized array are usually not more than a wavelength away from the origin (in $x$ and $y$ because they are constrained by the surface in $z$), which is why the performance reduces towards lower frequencies. 
For $N\gtrsim60$ and $f\gtrsim\SI{3}{\Hz}$, the mitigation factor is always larger than 5 with peak values $>20$, while a lower number of seismometers yields lower mitigation factors, for example $M>2$ for $f\gtrsim\SI{3}{\Hz}$ for the case of $N=20$.

The robustness of these arrays with respect to random displacement drawn from Gaussian distributions with $\sigma=\SI{50}{\m}$ is shown in the right panel of Figure~\ref{fig:BroadbandN}.
The central line shows the mean performance, while the band represents the standard deviation estimated from 10 different realizations of the displacements. The mitigation factor is most strongly reduced at high frequencies, while low frequencies are mostly unaffected. This is due to the different wavelengths of the seismic waves, where $\sigma=\SI{50}{\m}$ corresponds to $\lambda/8$ for S-waves at $f=\SI{10}{\Hz}$, while it is much smaller than the wavelength of P- and S-waves at $f=\SI{1}{\Hz}$.
Although the mitigation performance is reduced compared to the optimized positions, we observe more stability of the mitigation factor over the frequency range.
For $N\gtrsim60$, the mitigation factor is larger than 5 for $f\gtrsim\SI{3}{\Hz}$, even when considering the variations from different realizations of the random displacements.

As a next step, we optimize the positions of several seismometers per borehole.
We optimize the $x$- and $y$-positions of $N$ vertical boreholes at the same time as optimizing $z$-positions of $n$ seismometers inside each of these boreholes.

The broadband performance for $n=1,2,3,5$ and $10$ is shown in the left panel of Figure~\ref{fig:BroadbandMulti20} for $N=20$. Adding more seismometers in the boreholes strongly increases the mitigation factor. Although the positioning of the seismometers is constrained, they still provide significant additional information about the seismic field. With 5 seismometers in 20 boreholes, a similar performance can be achieved as with 60 boreholes with only a single seismometer per borehole. We also observe that the broadband performance is a bit less dependent on frequency than for the unconstrained optimization.
The mitigation factor is larger than 5 for $f\gtrsim\SI{3}{\Hz}$ for the case of 3 seismometers in each of the 20 boreholes.

\begin{figure}[h]
\makebox[\textwidth]{
    \centering    \subfloat{\includegraphics[width=.45\linewidth]{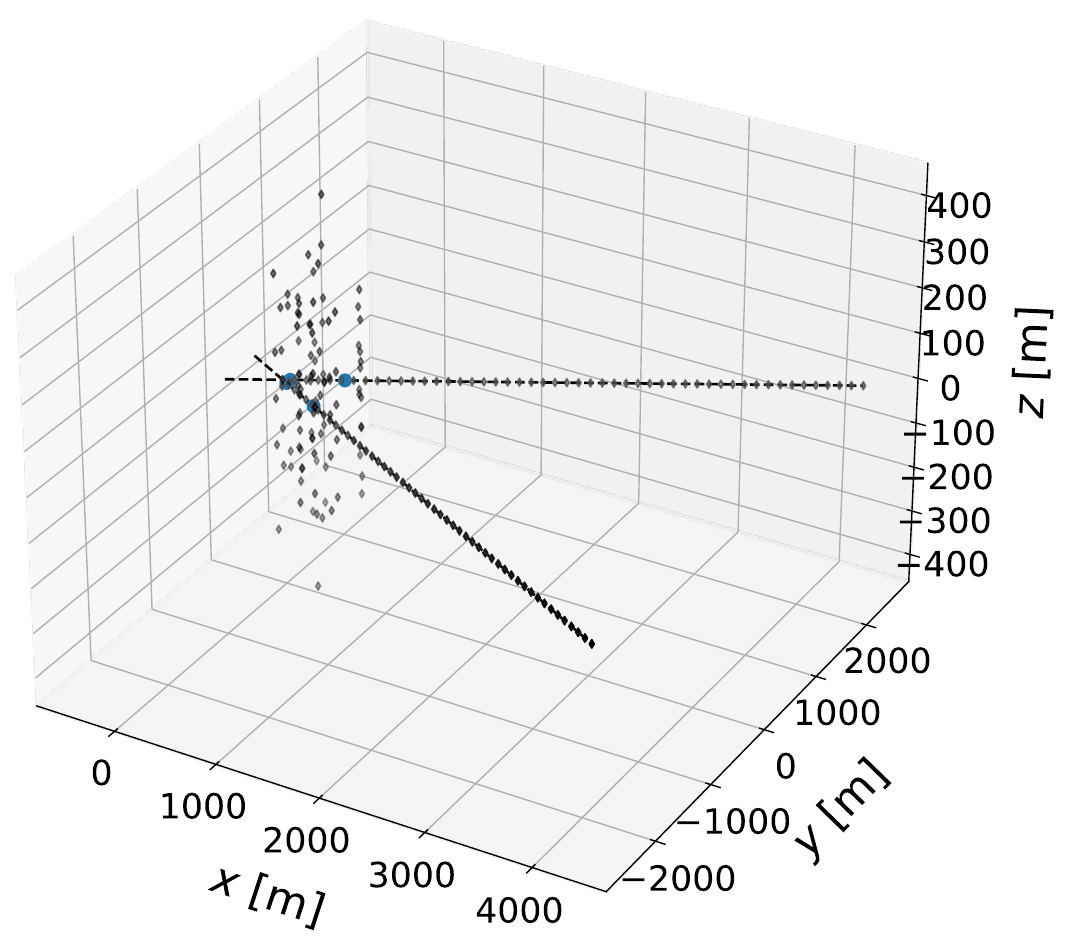}}\:\:
    \subfloat{\includegraphics[width=.45\linewidth]{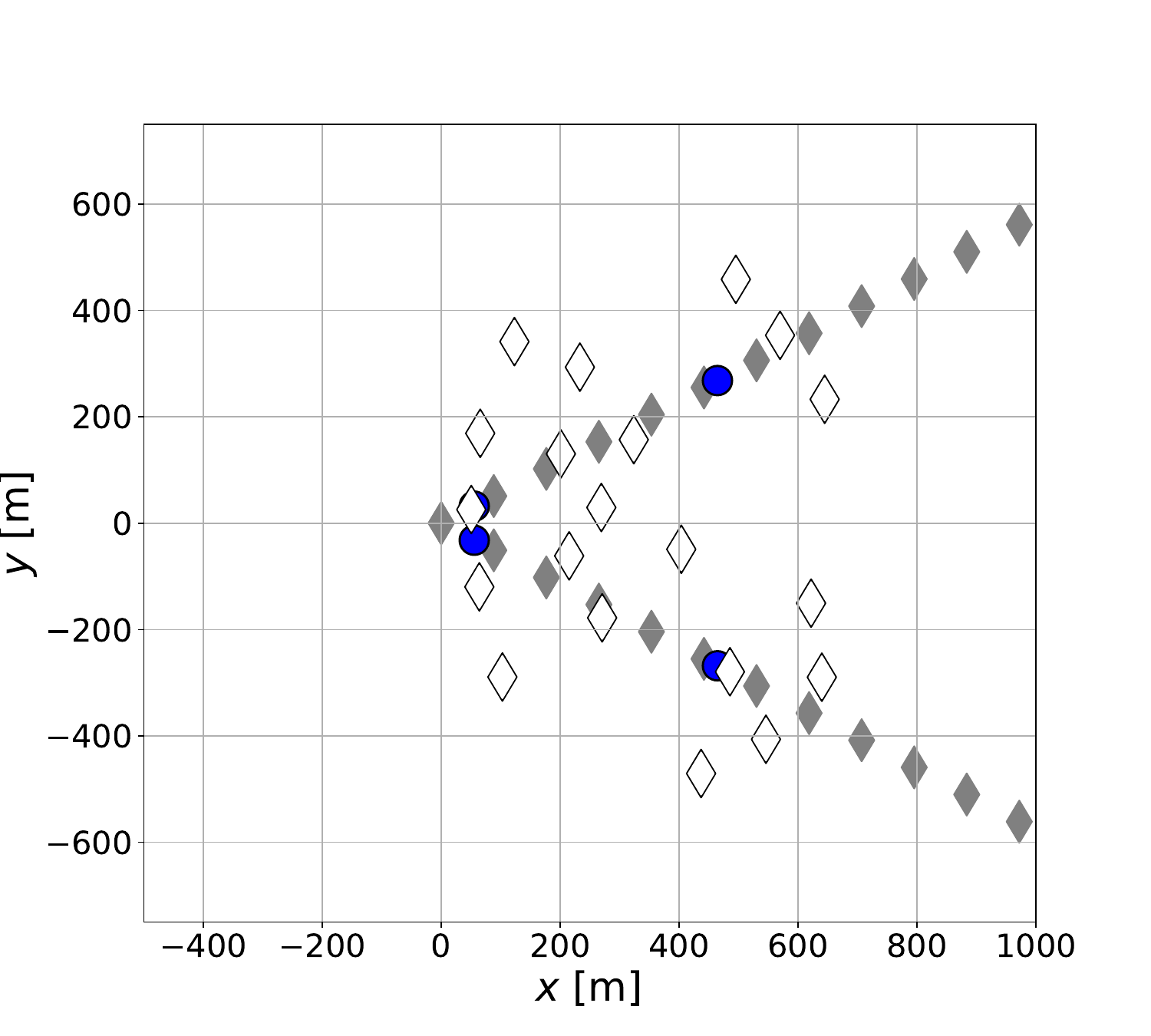}}
    }
    \caption{Left: Visualisation of the optimized array for 20 boreholes with 5 seismometers each, and the seismometers in the tunnels. Right: View from the top of the same array, zoomed into the area around the four mirrors (blue circles), with the positions of the boreholes shown with the open diamonds and the seismometers in the tunnels with the filled diamonds.}
    \label{fig:example_array}
\end{figure}

When testing the robustness of these arrays, we displace all seismometers in one boreholes together in $x$- and $y$-direction, i.e., we displace the whole borehole and do not consider horizontal displacements perpendicular to the borehole.
The seismometers are still independently displaced in $z$-direction, i.e., along the borehole.
The results are shown in the right side of Figure~\ref{fig:BroadbandMulti20}.
If there are at least $n=5$ seismometers in each borehole, the mitigation factor is $>5$ for $f\gtrsim\SI{3}{\Hz}$, similar to 60 boreholes with a single seismometer each. The mitigation factors for $n=3$ are only slightly worse.
As an example, the optimized array for $n=5$ is shown in Figure~\ref{fig:example_array}.

In addition, we test the effect of placing seismometers inside the ET infrastructure, in particular in the tunnels. As no additional boreholes need to be drilled, this is very cost-efficient. We test the gain of deploying 50 seismometers in the two tunnels that lead away from the corner for which we optimize. We use an even spacing of $\SI{100}{\m}$, so that we only consider seismometers up to half of the total length of each $\SI{10}{\km}$ tunnel of the triangle configuration. We choose to only study these fixed positions, because we found that optimizing the positions in the tunnels does not yield a large gain but does increase the complexity of the optimization task.


The results are shown in the left plot of Figure~\ref{fig:BroadbandMulti20T100}. The mitigation factor improves for all seismometer arrays by adding the seismometers in the tunnels. Already with $n=2$, mitigation factors of $\gtrsim5$ can be reached for $f\gtrsim\SI{4}{\Hz}$. With $n=10$, we even find $M>10$ for $f\gtrsim\SI{3}{\Hz}$, and peak performance of $M\sim34$.
The right plot of Figure~\ref{fig:BroadbandMulti20T100} shows the robustness with respect to displacements away from the optimized positions. 
In this configuration, we find particularly robust mitigation factors, so that for example for $n\geq3$, mitigation factors above 5 are obtained for all frequencies $f>\SI{3}{\Hz}$. For the best arrays, we find $M\gtrsim 10$ for $f\gtrsim\SI{3}{\Hz}$.

In a last step, we investigate the effect of increasing the number of boreholes to 50, combined with the effects of adding several seismometers per borehole as well as the seismometers in the interferometer tunnels.

In Figure~\ref{fig:BroadbandMulti50T100}, we show the broadband results for $N=50$ boreholes. Already $n=2$ seismometers per borehole are enough to achieve $M>10$ for $f\gtrsim\SI{4}{\Hz}$.
For $n=5$, we find $M > 10$ for $f\gtrsim\SI{3}{\Hz}$, reaching 
factors $M>40$ for $f\gtrsim\SI{8}{\Hz}$.
For $n=10$, we find $M>15$ for $f\gtrsim\SI{3}{\Hz}$, reaching $M>40$ for $f\gtrsim\SI{6}{\Hz}$ and peak factors of $M\sim60$.
We find that such a larger array is significantly more robust with respect to the random displacements than the array with $N=20$. Mitigation factors for $f\lesssim\SI{5}{\Hz}$ are almost unchanged.
Although the peak mitigations at larger frequencies drop with respect to the ideal positions, they never drop below the mitigation factors obtained for $f=\SI{3}{\Hz}$ and hence promising a mitigation performance that is stable over the frequency range.

\FloatBarrier

\section{Conclusions}
\label{sec:Conclusions}
We built on previous studies~\cite{FrancescaSingleMirrorOptimization,FrancescaJointMirrorOptimization,OurPaper}, and studied the broadband performance ($1-\SI{10}{\Hz}$) of seismometer array optimization for Newtonian noise mitigation at ET, as well as the robustness of the arrays with respect to random variations around the optimal seismometer positions. We tested the mitigation potential of deploying multiple seismometers per borehole, and supplementary seismometers in the interferometer tunnels as cost-effective solutions to expand an array for a given number of boreholes. We used arrays optimized for the largest frequency ($\SI{10}{\Hz}$) and studied benchmark cases of $N=20$ and $N=50$ boreholes with up to 10 seismometers per borehole. In addition, we studied the effect of adding $100$ additional seismometers in the tunnels without optimizing their positions.

We quantify the broadband performance in terms of mitigation factors, $M$, defined as the inverse of the maximum residuum for a full corner of ET, as well as the effect of random displacements.
We find that random displacements with a fixed length scale ($\SI{50}{\m}$) have a stronger impact on the mitigation performance at larger frequencies (smaller wavelengths), which leads to a reduced frequency dependence of the mitigation factors.
Adding multiple seismometers per boreholes achieves excellence mitigation factors, matching those of only slightly smaller arrays where a separate borehole would need to be drilled for each seismometer.
The additional seismometers in the interferometer tunnels further improve the mitigation potential.
In general, we find that larger arrays are more robust against displacements of the seismometers from their optimized positions.

In the case with the largest array that we studied ($N=50$ boreholes with $n=10$ seismometers per borehole and with seismometers in the tunnels), we find mitigation factors of $M>15$ for all frequencies $f\gtrsim \SI{3}{\Hz}$, even for the randomly-disturbed arrays. In a scenario with $N=20$ and $n=5$, we find $M>7$ for $f\gtrsim\SI{3}{\Hz}$ and $M>10$ for $f\gtrsim\SI{5}{\Hz}$, again for the randomly-disturbed arrays. Even in a scenario with only $n=3$ seismometers per borehole, we find $M>6$ for $f\gtrsim\SI{3}{\Hz}$ and $M>7$ for $f\gtrsim\SI{5}{\Hz}$ for the randomly-disturbed arrays.

We conclude that arrays that are optimized for large frequencies in the relevant frequency range are effective for broadband mitigation of Newtonian noise, although arrays optimized for broadband performance~\cite{FrancescaSingleMirrorOptimization} may  provide a better trade-off for the mitigation at different frequencies.
Our results on deploying arrays with multiple seismometers per borehole motivate research into the technical feasibility of such deployments as a cost-effective strategy for designing a seismic array for Newtonian noise mitigation.
We also find that equipping the ET infrastructure with seismometers promises to provide additional mitigation performance, inline with a study on fusion arrays~\cite{ophardt2025silencing}, and improves the robustness of the arrays with respect to random displacements from optimized seismometer positions.

Future research should test these conclusions in light of finite-elements simulations of seismic wave fields and the corresponding Newtonian noise~\cite{reumers2026efficientfiniteelementformulation,Schillings:2026ksw} in realistic geologies.
Moreover, it would be very interesting to expand these studies to fusion arrays of seismometers, strainmeters~\cite{ophardt2025silencing} and tiltmeters~\cite{DeSalvo:2026gla}, as well as surface arrays, and account for the possibility of directional drilling in the optimization.

\section*{Availability of Code}
The code used to produce the results in this paper is available at \url{https://github.com/lc316353/arrayOpt}. 

\section*{Acknowledgments}
This research was supported by the German Federal Ministry of Education and Research (BMBF) via project 05A2023 under grant number 05A23PA1.

\printbibliography

\end{document}